\documentclass[english,aps,amssymb,prl,twocolumn,showpacs]{revtex4}
\usepackage[T1]{fontenc}
\usepackage[latin9]{inputenc}
\usepackage{amsmath}
\usepackage{graphicx}
\usepackage{amssymb}
\usepackage{esint}

\makeatletter
\@ifundefined{textcolor}{}
{%
 \definecolor{BLACK}{gray}{0}
 \definecolor{WHITE}{gray}{1}
 \definecolor{RED}{rgb}{1,0,0}
 \definecolor{GREEN}{rgb}{0,1,0}
 \definecolor{BLUE}{rgb}{0,0,1}
 \definecolor{CYAN}{cmyk}{1,0,0,0}
 \definecolor{MAGENTA}{cmyk}{0,1,0,0}
 \definecolor{YELLOW}{cmyk}{0,0,1,0}
 }

\renewcommand{\phi}{\varphi}
\renewcommand{\epsilon}{\varepsilon}

\renewcommand{\vec}[1]{{\bf #1}}

\usepackage{epsfig}
\@ifundefined{showcaptionsetup}{}{%
\PassOptionsToPackage{caption=false}{subfig}}
\usepackage{subfig}
\makeatother

\usepackage{babel}

\begin{document}

\title {Magnetoelectric effects in superconducting nanowires with Rashba spin-orbit coupling}

\author{Teemu Ojanen$^{1,2}$}
\email[Correspondence to ]{teemuo@boojum.hut.fi}
\affiliation{$^1$Low Temperature Laboratory, Aalto University, P.~O.~Box 15100,
FI-00076 AALTO, Finland }
\affiliation{$^2$Physics Department, Harvard University, Cambridge, Massachusetts 02138, USA}
\date{\today}
\begin{abstract}
Recent experiments in semiconductor nanowires with a spin-orbit coupling and proximity-induced superconductivity exhibit signatures of Majorana bound states predicted to exist in the topological phase. In this work we predict that these nanowire systems exhibit unconventional magnetoelectric effects showing a sharp crossover behavior at the topological phase transition. We find that magnetic fields with a component parallel to the spin-orbit field can give rise to currents in equilibrium. Surprisingly, also fields perpendicular to the spin-orbit field may induce currents and can be employed in adiabatic charge pumping. The perpendicular field magnetoelectric effect may be regarded as a manifestation of the anomalous Hall effect in one dimension. We discuss how the predicted phenomena could be observed in experiments and employed in probing the topological phase transition.

\end{abstract}
\pacs{73.63.Nm, 74.78.Na,74.78.Fk,}
\maketitle
\bigskip{}

\emph{Introduction}-- Recent experiments in nanowires with a strong spin-orbit coupling and proximity superconductivity show intriguing signatures of Majorana bound states \cite{mourik,das,deng}, a particle-like many-body excitations behaving as their own antiparticles. The surge of experimental and theoretical activity around the topic was largely initiated by the prediction that these systems exhibit topological superconductivity \cite{lutchyn1,oreg}. Shortly before it was proposed that two-dimensional magnetic Rashba systems in the proximity of a superconductor provided a candidate to realize topological superconductivity \cite{lutchyn,sato}. However, tailoring suitable systems and controlling relevant physical parameters have proven more feasible in nanowires \cite{lutchyn1,oreg,alicea1}. Nanowire systems have also opened up exciting possibilities to experimentally study effects resulting from the three necessary ingredients of the topological phase, a spin-orbit coupling, superconductivity and magnetic fields.

In this work we predict and study properties of magnetoelectric effects in 1d wires resulting from the interplay of the Rashba coupling, superconductivity and magnetic fields. These magnetoelectric effects lead to non-dissipative electric currents that are not related to the phase gradient of the superconducting order parameter. The magnetoelectric effects exhibit a clear crossover behavior when the nanowire undergoes a topological phase transition, thus providing a method to probe the transition. We study two qualitatively different effects distinguished by whether the applied magnetic field has a finite component parallel to the spin-orbit field or not. Experiments have demonstrated that the direction of the spin-orbit field can be resolved quite accurately \cite{mourik, nadj}.
\begin{figure}[t]
\centering
\includegraphics[width=0.4\columnwidth, clip=true]{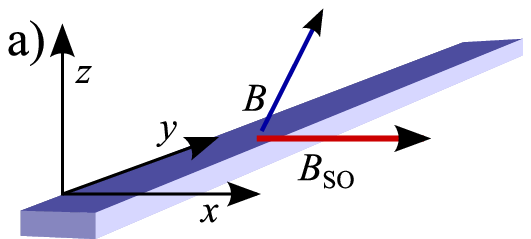}
\includegraphics[width=0.4\columnwidth, clip=true]{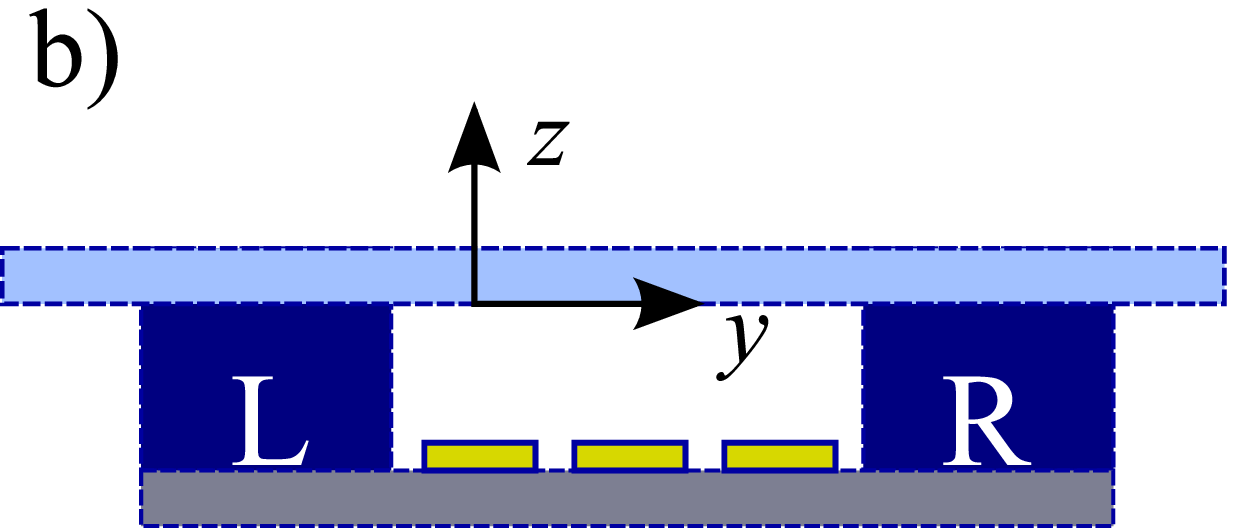}
\caption{a):  Superconducting nanowire with a Rashba spin-orbit coupling in external magnetic field $B$ where $B_{SO}$ denotes the spin-orbit field for particles moving in the positive $y$ direction. b): Experimental setup for probing the magnetoelectric effects.}
\label{sceme}
\end{figure}

First we predict that magnetic fields having a finite component parallel to the spin-orbit field lead to a finite DC current through the wire. This effect is sensitive to the closing of the energy gap associated to the topological phase transition. Then we show that electric currents can also be driven by magnetic fields perpendicular to the spin-orbit field. This phenomenon arises from the same band curvature effect that contributes to the anomalous Hall effect in 2d Rashba systems and enables adiabatic charge pumping when magnetic field is periodically modulated. We discuss how the predicted magnetoelectric effects could be observed in experiments.

\emph{Model system}--
We consider a nanowire with a Rashba spin-orbit coupling in the proximity of a superconductor in magnetic field, depicted in Fig.~\ref{sceme}.
The wire is described by the Bogoliubov-de Gennes Hamiltonian \cite{lutchyn1, oreg},
\begin{align}\label{h}
H(k)=&\left(\epsilon_k+\alpha k\sigma_x\right)\tau_z+\boldsymbol{B}\cdot\vec{\sigma} +\Delta\,\tau_x,
\end{align}
where $\epsilon_k=\frac{\hbar^2k^2}{2m}-\mu$ and $\sigma_i$ and $\tau_i$ are Pauli matrices operating in the spin and the particle-hole space, respectively. Hamiltonian (\ref{h}) is written in the Nambu basis  $\Psi=(\psi_{k\uparrow}, \psi_{k\downarrow},\psi^{\dagger}_{-k\downarrow},-\psi^{\dagger}_{-k\uparrow} )^T$ and the coordinates have been chosen so that $k$ is the momentum along the $y$-axis and the effective spin-orbit field is parallel to $x$ direction. The first term corresponds to kinetic energy of electrons and holes including the Rashba coupling characterized by a coupling constant $\alpha$. The second term is the Zeeman coupling $\vec{B}=(B_x,B_y,B_z)$ due to the applied magnetic field and the last term arises from the proximity-induced superconducting pairing. Hamiltonian (\ref{h}) possesses a particle-hole symmetry and the four energy bands satisfy $E_{-i}(\vec{k}) = -E_{i}(\vec{-k})$ for $i=1,2$ where the corresponding eigenstates $i$ and $-i$ are related by a particle-hole transformation. In the case $B_x=0$, the spectrum admits an analytical solution
\begin{align}\label{E}
E_{1/2}^2(\vec{k})=\epsilon_k^2&+\alpha^2k^2+B^2+\Delta^2\nonumber \\ &\mp2\sqrt{B^2(\epsilon_k^2+\Delta^2)+\epsilon_k^2\alpha^2k^2},
\end{align}
where $B^2=B_y^2+B_z^2$. We adopt a convention according which $B_x$ (that is parallel to the effective spin-orbit field) is called \emph{parallel} magnetic field and $B_y$ and $B_z$ are called \emph{perpendicular} fields.
For the reference, the relevant physical parameters in experiment employing InSb wires in Ref.~\cite{mourik} were $m=0.015\,m_e$,  $E_R=\frac{2m\alpha^2}{\hbar^2}=200$ $\mu$eV, $\Delta=250$ $\mu$eV (at T=60 mK) and $|\vec{B}|\lesssim 1.5$ meV (for fields < 1T). The transverse mode separation in the wire was several meV which is much larger than the other energy scales.

\emph{Magnetoelectric effect for parallel fields $B_x\neq0$}--
In this section we consider magnetic fields that have a non-vanishing parallel component with the spin-orbit field (according to conventions of Fig.~\ref{sceme} this means that $B_x\neq 0$). This leads to a non-vanishing current through the wire, an effect which is similar to the magnetoelectric effect predicted by Edelstein in 2d systems \cite{edelstein,yip}.

Current through the wire can be evaluated as
\begin{equation}\label{cu1}
J=\frac{1}{L}\sum_{k,i} \langle n_i|\widehat{J}|n_i\rangle f_i =\frac{1}{L}\sum_{k,i}\mathrm{Tr}\left[ P_i \widehat{J}\right]f_i ,
\end{equation}
where  $\widehat{J}=\frac{e}{2\hbar}(\frac{\hbar^2 k}{m}+\alpha\sigma_x)=\frac{e}{2\hbar}\partial_kH\tau_z$ is the current operator, $P_i$ is a projection operator to the eigenstate $|n_i\rangle$ of Hamiltonian (\ref{h}) with energy $E_i$ and $f_i$ is the Fermi distribution at energy $E_i$. The projection operators are given by $P_i=\prod_{j\neq i}\frac{H-E_j}{E_i-E_j}$ for the four bands and the trace should be calculated over the Nambu and spin indices. A further evaluation of Eq.~(\ref{cu1}) yields
$J=\frac{e}{2}\int\frac{dk}{2\pi}\sum_{i}\left( \frac{\hbar k}{m}+\frac{\alpha}{\hbar}\mathrm{Tr}\left[ P_i \sigma_x\right]\right)f_i.$
The general formula for the current in terms of the energies is lengthy and presented in the supplement. Current is an odd function of the parallel field $J(B_x)=-J(-B_x)$, a property which can be employed in distinguishing the magnetoelectric contribution. In the linear order in $B_x$ current is given by
\begin{align}\label{cu4}
J=&\frac{e\alpha B_x}{\hbar}\int\frac{dk}{2\pi}\sum_{i=1,2}\frac{2(1-2f_i)(-1)^i}{E_iD}\left[ \left((-1)^i-\frac{4E_i^2}{D} \right)\times\right.\nonumber\\ &\frac{4\epsilon_k^2\alpha^2k^2}{D}\left. +\left(\Delta^2+\epsilon_k^2+\alpha^2k^2+(-1)^{i}\frac{D}{4}\right) \right],
\end{align}
where $D=4\sqrt{B^2(\epsilon_k^2+\Delta^2)+\epsilon_k^2\alpha^2k^2}$. The wire undergoes a topological phase transition when the perpendicular field satisfies $B=\sqrt{\mu^2+\Delta^2}$ \cite{oreg,lutchyn1}, associated with the energy gap closing $E_{\pm 1}(k)=0$ at $k=0$. The numerator in the integrand for $i=1$ term also vanishes at that point and current remains analytic.
\begin{figure}[t]
\centering
\includegraphics[width=0.45\columnwidth, clip=true]{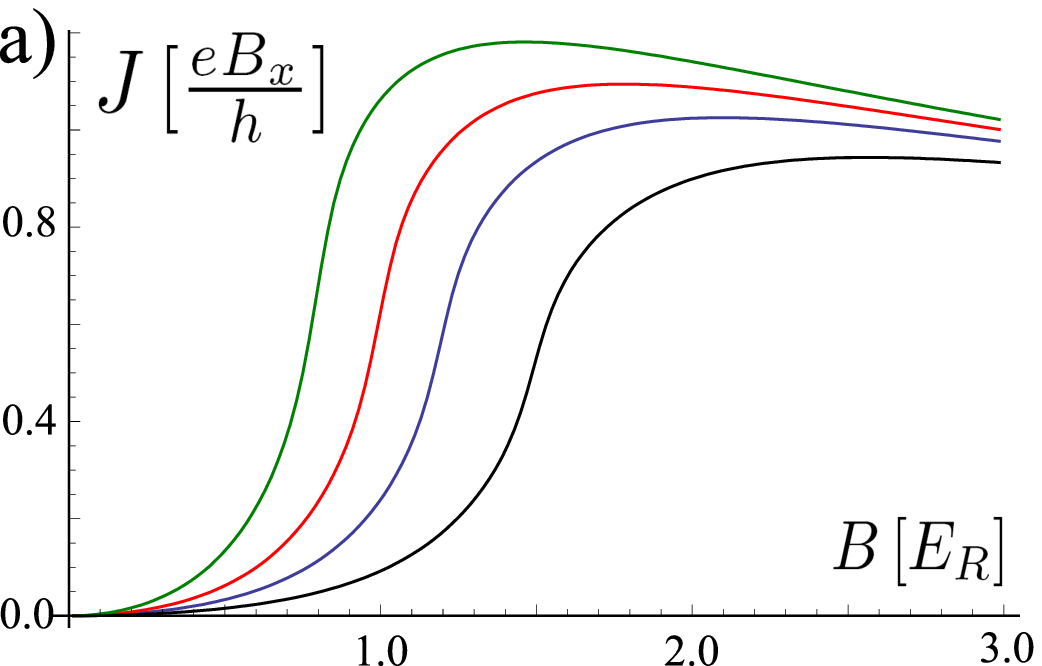}
\includegraphics[width=0.45\columnwidth]{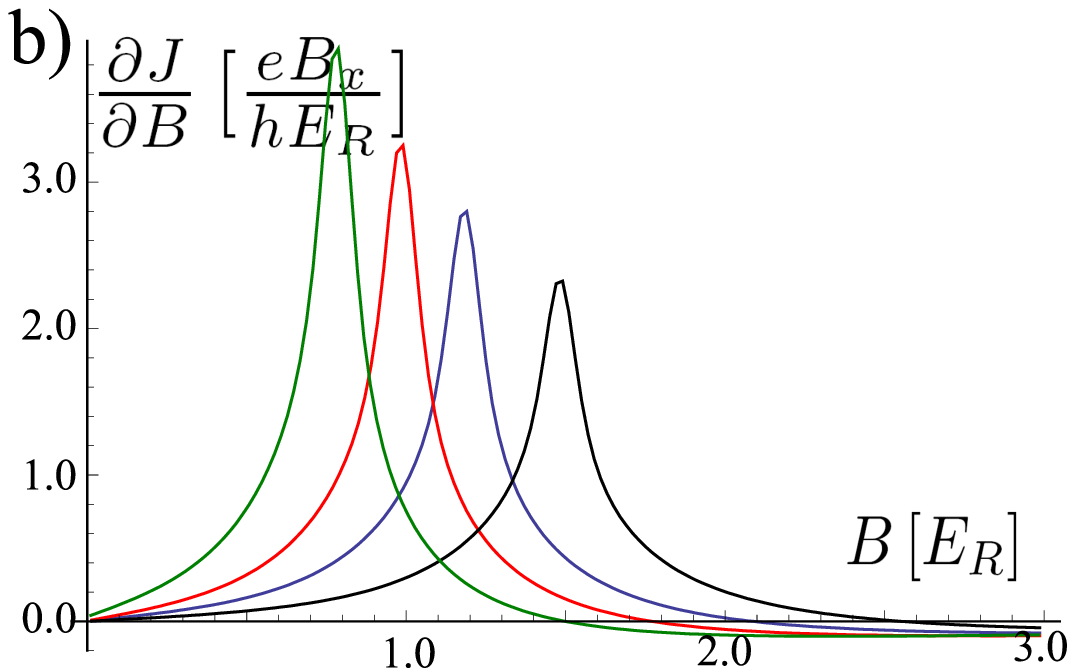}
\caption{  a):  Current from Eq.~(\ref{cu4}) a function of the perpendicular field $B$ for $T=0.025\,E_R$, $\mu=0.0$, and  $\Delta=0.8, 1.0, 1.2, 1.5\times E_R$ (top to bottom). b): Differential current $\frac{\partial J}{\partial B}$  calculated from the curves in a). The maxima coincide with the critical field above which the system is in the topological phase.}
\label{spect}
\end{figure}

Current (\ref{cu4}) is plotted as a function of the perpendicular field $B$ in Fig.~\ref{spect} a). Current starts from zero at $B=0$ and increases rapidly until saturating and starting to decrease slowly. Due to the combined effect of the spin-orbit coupling and magnetic field $B_x$, the dispersion becomes asymmetric $E_i(k)\neq E_i(-k)$. This asymmetry and superconductivity are necessary ingredients to achieve finite current since in the normal state the current cancels \footnote{In the normal state the cancelation follows from $J=\int dk\partial_kE=E(k_{F+})-E(k_{F-})=E_F-E_F=0$, where $E(k)$ is the normal state spectrum. In the superconducting state current is \emph{not} related to the group velocity of the Bogoliubov spectrum.}. For large $B$, the eigenstates resemble magnetic rather than helical metal. For a dominantly magnetic dispersion $B_x$ induced  $k\longleftrightarrow -k$ spectral asymmetry becomes suppressed, thus explaining the saturation and decrease of current for large $B$. The saturation sets on when $\partial_B J$ achieves maximum which, as illustrated in Fig.~\ref{spect} b), coincides accurately with the topological phase transition. The results in Fig.~\ref{spect} are calculated at temperature which translates to 60 mK for parameters of Ref.~\cite{mourik}. The peaks in Fig.~\ref{spect} b) get broadened and shorter as the temperature increases, but the well-defined structure survives to well above $T=0.1\,E_R$. Thus the parallel magnetoelectric effect provides a method to probe the phase transition indirectly. It is important to find alternative methods to characterize the phase transition since there exists theoretical evidence that some of the considered signatures of the topological phase, such as the zero bias peak in the tunneling current \cite{kells} and the fractional Josephson effect \cite{sau}, could have explanations different from Majorana bound states.

So far we have analyzed translation-invariant systems but experiments are performed with finite wires. For experimental implications we need to consider a setup in Fig.~\ref{sceme} b) where a finite wire is laterally coupled to two reservoirs L and R. Superconductivity in the wire is induced either by one or both of the reservoirs. Both of the reservoirs need to be superconducting in order to achieve finite current since equilibrium currents cannot be sustained in a normal metal reservoir. In practice the wire may support multiple transverse modes as in Ref.~\cite{mourik}. However, when the mode separation is large compared to the other energy scales, as in \cite{mourik}, modes can be thought as independent (at least in the clean limit). In addition, the modes below the topmost one do not contribute to current when the mode separation is much larger than other characteristic energy scales, so the single mode approximation is expected to be an accurate starting point.

It is also interesting to consider the case where one of the reservoirs is in the normal state or the wire is effectively pinched of by a local gate so that current is suppressed. According to thermodynamic arguments considered in the 2d case \cite{yip}, the magnetic field induced current is canceled by a superconducting phase gradient along the wire in equilibrium. Thus the effect could be perhaps also observed by $B_x$-dependent build up of superconducting phase when current is blocked.

\emph{Magnetoelectric effect for perpendicular fields: linear response }--
Now we consider magnetic fields that are perpendicular to the spin-orbit field which in our conventions mean that $\vec{B}$ lies in the $y-z$ plane ($B_x=0$).  Below we derive the surprising result that perpendicular fields can also be employed to drive currets in the wire. This magnetoelectric effect arises from essentially the same band curvature effect that gives rise to the anomalous Hall effect in 2d Rashba systems with perpendicular magnetization \cite{nagaosa,ojanen}.

\begin{figure}[t]
\centering
\includegraphics[width=0.4\columnwidth, clip=true]{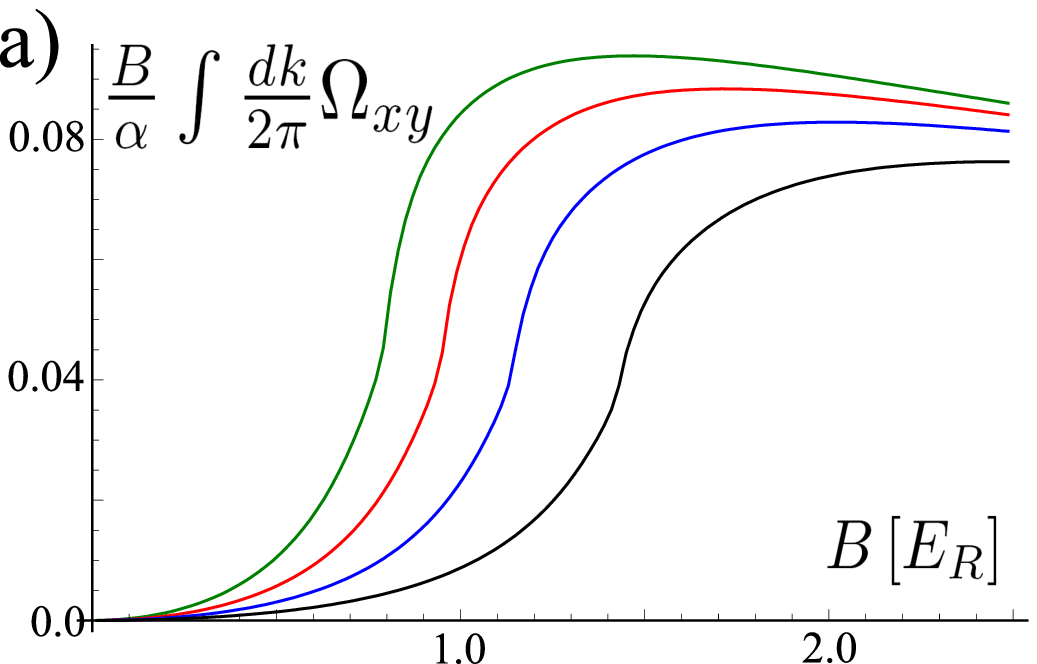}
\includegraphics[width=0.35\columnwidth, clip=true]{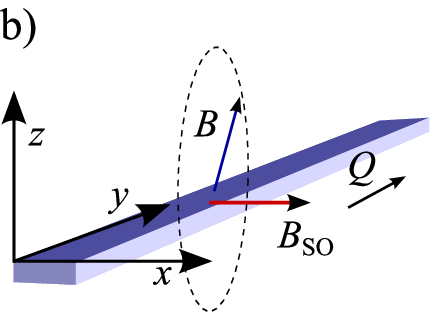}
\caption{  a):  Magnitude of the perpendicular effect as a function of the field $B$ for $T=0.025\,E_R$, $\mu=0.0$ and  $\Delta=0.8, 1.0, 1.2, 1.5\times E_R$ (from top to bottom). b): Adiabatic pumping scheme. When the perpendicular field $B$ traverses a loop in the $y-z$ plane, charge $Q$ is pushed through a cross section.}
\label{om}
\end{figure}

First we consider a magnetic field configuration with a static component $B$ in $z$-direction and a weak time-dependent component $\delta B_y$ in $y$-direction. Treating $\delta B_y$ as a perturbation, current in the wire is given by the standard linear response theory as
\begin{equation}\label{lin1}
J(\omega)_\perp=\chi(\omega)\delta B_y(\omega).
\end{equation}
The imaginary-time representation of the response function $\chi(\omega)$ is given by
\begin{align}\label{lin2}
\chi(i\omega_m)=\frac{1}{L\beta}\sum_{k,n}\mathrm{Tr}[\widehat{J}G(i(\omega_m+\nu_n))\sigma_yG(i\nu_n)],
\end{align}
where $\beta$ is inverse temperature and the Matsubara Green's functions $G(i\nu_n)$ are defined by $G(i\nu_n)=\frac{1}{i\nu_n-H}$, where $H$ is given by Eq.~(\ref{h}) with $\vec{B}=(0,0,B)$. Expression (\ref{lin2}) can be evaluated using methods of Ref.~\cite{ojanen} where similar expressions were evaluated in 2d geometry.
 In the following we are interested in low-frequency dynamics. In $\omega\to 0$ limit $\chi(\omega)$ approaches to zero so it is convenient to consider the quantity  $\lim_{\omega\to 0}\chi(\omega)/i\omega\equiv\frac{e}{\alpha} \int  \frac{dk}{2\pi}\Omega_{xy}$, where
\begin{align}\label{omeg}
\Omega_{xy}(k)=\left(-\frac{i\alpha^2 }{2}\right)\sum_{n,n'}\frac{\mathrm{Tr}\left[\sigma_x P_{n'}\sigma_y P_n\right] }{(E_n-E_{n'})^2}(f_n-f_{n'}).
\end{align}
Here $P_n$ is a projection operator to the eigenstate with eigenvalue $E_n$ given by Eq.~(\ref{E}). The explicit form of $\Omega_{xy}(k)$ is long and given in the supplement. Current in the wire can now be expressed as
\begin{equation}\label{lin3}
J(t)=e\frac{\partial_t\delta B_y(t)}{B}\left(\frac{B}{\alpha}\int \frac{dk}{2\pi}\Omega_{xy}(k)\right).
\end{equation}
Thus we have reached a remarkable conclusion that temporal variations of magnetic field parallel to the wire can drive current. Figure \ref{om} a) illustrates the behavior of dimensionless quantity $\frac{B}{\alpha}\int\frac{dk}{2\pi}\Omega_{xy}$ setting the magnitude of the effect. At low temperatures it behaves essentially as the current from the parallel magnetoelectric effect plotted in Fig.~(\ref{spect}) a), therefore the differential current $\partial_B J$ resulting from the perpendicular effect (\ref{lin3}) exhibits peaks similar to those shown in Fig.~(\ref{spect}) b). From symmetry considerations it follows that result (\ref{lin3}) holds even if the directions of the static and dynamic magnetic fields $B$ and $\delta B_y$ are rotated arbitrarily around the $x$-axis as far as they remain perpendicular to each other. We are not aware of previous predictions of constitutive relations between current and magnetic field of type (\ref{lin3}) in 1d wires.

The magnetoelectric effect (\ref{lin3}) can be understood as a manifestation of the same band curvature effect that gives rise to the anomalous Hall effect in 2d Rashba systems \cite{nagaosa}. Recently the anomalous Hall conductivity was calculated in superconducting two-dimensional Rashba systems with magnetization perpendicular to plane \cite{ojanen}. The Hall conductivity is given by $\sigma_{xy}=\frac{e^2 }{\hbar}\int\frac{d^2k}{(2\pi)^2}\Omega_{xy}(k)$ where momentum argument should be interpreted as the magnitude of the in-plane momentum $k=\sqrt{k_x^2+k_y^2}$ and the static part of the magnetic field $B$ corresponds to magnetization perpendicular to the plane. Therefore $\Omega_{xy}(k)$ can be regarded as the superconducting counterpart of the Berry curvature in non-superconducting system \cite{nagaosa,xiao}. The reason why Eq.~(\ref{lin3}) is connected to the 2d Hall conductivity can be traced to the formal similarity of response function (\ref{lin2}) and the off-diagonal current correlation function in 2d magnetic Rashba system \footnote{Intuitively, the connection between the response function (\ref{lin2}) and the Hall conductivity can be thought of as "replacing" one current direction by a spin direction. After all, the current operator contains a spin part.}.


The magnetoelectric effect (\ref{lin3}) could be observed in a setup of Fig.~\ref{sceme} b) where both reservoirs are superconducting. For example, by applying a static field and a small AC field perpendicular it, both lying in the $y-z$ plane, the wire should exhibit a measurable AC current. Current as a function of the static field also exhibits a clear crossover behavior at the topological phase transition and could be employed to probe it.

\emph{Magnetoelectric effect for perpendicular fields: adiabatic pumping}-- The relation between current and perpendicular magnetic field (\ref{lin3}) suggests that the integrated charge flowing through a cross section when the field is modified depends only on the net change of the field but not on time that the procedure lasts. This is a hallmark of adiabatic transport and geometric pumping \cite{thouless, xiao}. Below we discuss pumping processes where magnetic field executes a cycle in the $y-z$ plane ($B_x=0$ throughout this section), as illustrated in a particular case in Fig.~\ref{om} b). The calculation proceeds along the lines of adiabatic transport and polarization of insulators \cite{xiao}.

Suppose that some  parameter $\lambda(t)$ in Hamiltonian (\ref{h}) varies slowly in time $H(t)=H(\lambda(t))$. Let us further assume that the system is in the ground state and that the population of the energy band $E_n$ is $f_n$ (for negative energy bands $f_n=1$ and for the positive energy bands $f_n=0$) initially at $t=0$.  Current in the wire for $t>0$ is given by
\begin{equation}\label{cur1}
J(t)=\int\frac{dk}{2\pi}\sum_{n}\langle\psi_n(t)|\widehat{J}|\psi_n(t)\rangle f_n,
\end{equation}
where again $\hat{J}=\frac{e}{2\hbar}\partial_kH\tau_z$. Initially the states $|\psi_n(t)\rangle$  satisfy $|\psi_n(0)\rangle=|n(0)\rangle$ where $|n(0)\rangle$ is an eigenstates of (\ref{h}) at $t=0$. According to the adiabatic theorem, the temporal evolution of an eigenstate is given by $|\psi_n(t)\rangle=e^{-i\int_0^t E_n(s)ds}\left( |n(t)\rangle-i\hbar\sum_{n'\neq n}\frac{|n'(t)\rangle\langle n'(t)|\partial_t n\rangle}{E_n-E_{n'}} \right)$ in the lowest order in the time derivatives. Here $|n(t)\rangle$ and $E_{n}$ correspond to the instantaneous eigenstates and eigenvalues of the time-dependent Hamiltonian. Following the the standard steps in adiabatic transport theory, when the adiabatic parameter traces path from $\lambda(0)=\lambda_i $ through $\lambda(T_f)=\lambda_f$, the total charge $Q=\int_{0}^{T_f}dt  J (t)$ pumped through the wire is
\begin{align}\label{curx}
Q=&\left(-\frac{ie}{2}\right)\int_{\lambda_i}^{\lambda_f}d\lambda \int\frac{dk}{2\pi}\nonumber\\
&\sum_{n,n'(n'\neq n)}\frac{\mathrm{Tr}\left[\partial_kH\tau_z P_{n'}\partial_\lambda H P_n\right] }{(E_n-E_{n'})^2}(f_n-f_{n'}).
\end{align}
Here we have introduced projection operators $P_n(\lambda)$ to instantaneous eigenstates corresponding to $E_n(\lambda)$. The integrand in Eq.~(\ref{curx}) is analogous to the Berry curvature in non-superconducting systems \cite{xiao}. The extra Nambu matrix $\tau_z$ inside the trace, arising from the Nambu structure of the current operator, makes the analogy to the Berry curvature incomplete and the pumped charge is not quantized. Result (\ref{curx}) can be generalized to finite temperatures by replacing $f_n$ by Fermi distribution at energy $E_n(\lambda=0)$ if energies $E_n$ do not change during the process or the cycle is faster than relaxation processes  (while being slow compared to the relevant spectral gaps as the adiabatic approximation requires).

Let us now consider the pumped charge when magnetic field is modified slowly in the $y-z$ plane. The cycle executed by the magnetic field is conveniently parameterized by magnitude $B(\lambda)$ and angle $\phi(\lambda)$ so that field-dependent part of Eq.~(\ref{h}) becomes $\vec{B}\cdot \sigma=B\,\mathrm{cos}(\phi)\sigma_y+B\,\mathrm{sin}(\phi)\,\sigma_z$. Application of Eq.~(\ref{curx}) yields
\begin{align}\label{q1}
Q=\left(\frac{e}{\alpha}\right)\int_{\lambda_i}^{\lambda_f}d\lambda B\partial_\lambda\phi\int\frac{dk}{2\pi}\Omega_{xy}(k),
\end{align}
were $\Omega_{xy}(k)$ is given by Eq.~(\ref{omeg}). For a general cycle $(B(\lambda),\phi(\lambda))$ also $\Omega_{xy}(k)$ depends on $\lambda$ through $B$ (see the supplement).
Expression (\ref{q1}) is the general formula for the adiabatically pumped current when magnetic field undergoes a cycle in $y-z$ plane. To be more concrete, we consider a cycle where the magnitude of the magnetic field is kept constant $B(\lambda)=B$ but the angle makes a full rotation $\phi(\lambda_f)-\phi(\lambda_i)=2\pi$. Then $\Omega_{xy}(k)$ is independent of $\lambda$ and  the pumped charge in the cycle becomes
\begin{align}\label{q2}
Q=2\pi e\left(\frac{B}{\alpha} \int\frac{dk}{2\pi}\Omega_{xy}(k)\right).
\end{align}
The sign of $Q$ is inverted when the field executes the loop in the reversed direction. Pumping formulas (\ref{q1}), (\ref{q2}) remain valid in the vicinity of the topological phase transition despite the gap $E_1(k=0)-E_{-1}(k=0)$ may become arbitrarily small because a field-induced mixing of the $E_{-1}(k)$ and $E_1(k)$ bands is prohibited (as pointed out in the supplement). Validity of pumping results requires executing the cycle in time $T$ which satisfies $T^{-1}\ll \min_k (E_2(k)-E_1(k))/\hbar$. The results in Fig.~\ref{om} a) show that for the chosen parameter values the pumped charge per cycle is of the order of $Q\sim 0.5\, e$ in the topological phase.

It was recently demonstrated experimentally that the direction of the spin-orbit field can be resolved quite accurately and indeed is perpendicular to the wire \cite{mourik, nadj}. Considering that in these experiments magnetic field was also rotated in the plane perpendicular to the spin-orbit field, central ingredients to test prediction (\ref{q2}) experimentally seems to be realized. Finite size effects and disorder could modify the quantitative results, but in the light of recent developments an experimental realization of adiabatic pumping (\ref{q2}) seems promising.

\emph{Conclusion}-- In this paper we predicted two distinct magnetoelectric effects in superconducting nanowires with Rashba coupling, giving rise to non-dissipative currents as a response to applied magnetic fields. The two effects, distinguished by whether the applied field has a parallel component with the spin-orbit field or not, result from the interplay of the Rashba coupling, magnetic fields and superconductivity. We proposed that the predicted effects could be observed in recently realized experimental setups and employed in probing the topological phase transition and realizing adiabatic charge pumping.

It is pleasure to thank Jay Sau and Takuya Kitagawa for discussions. The author acknowledges Academy of Finland for support.

\widetext

\section{Supplementary material}
Here we present supplementary results and formulas that were cited in the main text.

\subsection{Magnetoelectric effect for parallel fields $B_x\neq0$}
In the main text it was noted that application of magnetic fields with a non-vanishing  $B_x$ component (which is, according to our conventions, \emph{parallel} to the effective magnetic field arising from the spin-orbit coupling) give rise to current
\begin{equation}\label{cuapp}
J=\frac{e}{2}\int\frac{dk}{2\pi}\sum_{i}\left( \frac{\hbar k}{m}+\frac{\alpha}{\hbar}\mathrm{Tr}\left[ P_i \sigma_x\right]\right)f_i\equiv J_0+J_1,
\end{equation}
where $P_i=\prod_{j\neq i}\frac{H-E_j}{E_i-E_j}$ are the projection operators to the four eigenstates  $E_i$ of the Hamiltonian (\ref{h}) of the main text and the trace should be calculated over the Nambu and spin indices. Current components $J_0$ and $J_1$ correspond to the first and the second term in the curly brackets in Eq.~(\ref{cuapp}). Due to the finite parallel component $B_x\neq0$, the energy bands do not admit analytical expressions. The explicit formulas for the traces in terms of $E_i$ are given by
\begin{equation}\label{tr1}
\mathrm{Tr}\left[ P_i \sigma_x\right]=\frac{4}{\prod_{j\neq i}(E_i-E_j) } \left[B_x^3+2\epsilon_k\alpha kE_i+ B_x\left(3\Delta^2+3\epsilon_k^2+3\alpha^2k^2+B^2+E_i^2-\frac{1}{2}\sum_{k}E_k^2  \right)   \right]
\end{equation}
so $J_1$ becomes
\begin{equation}\label{cu3}
J_1=\frac{e\alpha}{2\hbar}\int\frac{dk}{2\pi}\sum_{i}\frac{4f_i}{\prod_{j\neq i}(E_i-E_j) }\left[B_x^3+B_x\left(3\Delta^2+3\epsilon_k^2+3\alpha^2k^2+B^2+E_i^2-\frac{1}{2}\sum_{k}E_k^2 \right) +2\epsilon_k\alpha kE_i   \right].
\end{equation}
From the property $E_i(k,B_x)=E_i(-k,-B_x)$, it follows that current $(\ref{cuapp})$ is an odd function of the parallel field $J(B_x)=-J(-B_x)$ (implying that  $J(0)=0$). Calculation of current generally requires solving the energy bands numerically and then evaluating $J_0$ and $J_1$.
However, to evaluate current in the linear order in $B_x$ one can make further progress analytically. By applying the first order perturbation theory, the energy bands can be solved as $E_{\pm1}(B_x)=\pm E_{1}(B_x=0)-\delta E$ and $E_{\pm 2}(B_x)=\pm E_{2}(B_x=0)+\delta E$ where $\delta E=\frac{4B_x\epsilon_k\alpha k}{E_2^2-E_1^2}$ and $E_{i}(B_x=0)$ are given by Eq.~(\ref{E}) in the main text. Current up to the linear order in $B_x$ can then be evaluated by inserting the expressions for energies in $(\ref{cu3})$ and retaining only linear terms. This procedure yields
\begin{align}\label{acu4}
J_1=&\frac{e\alpha B_x}{2\hbar}\int\frac{dk}{2\pi}\sum_{i=-2}^{2}\frac{4f_i(-1)^i}{E_i(E_2^2-E_1^2)}\left[ \left((-1)^i-\frac{4E_i^2}{E_2^2-E_1^2} \right)\frac{4\epsilon_k^2\alpha^2k^2}{E_2^2-E_1^2}\right.\left. +\left(\Delta^2+\epsilon_k^2+\alpha^2k^2+(-1)^{i}\frac{E_2^2-E_1^2}{4}\right) \right]
\nonumber\\ =&\frac{e\alpha B_x}{2\hbar}\int\frac{dk}{2\pi}\sum_{i=1,2}\frac{4(1-2f_i)(-1)^i}{E_i(E_2^2-E_1^2)}\left[ \left((-1)^i-\frac{4E_i^2}{E_2^2-E_1^2} \right)\frac{4\epsilon_k^2\alpha^2k^2}{E_2^2-E_1^2}\right. \left. +\left(\Delta^2+\epsilon_k^2+\alpha^2k^2+(-1)^{i}\frac{E_2^2-E_1^2}{4}\right) \right].
\end{align}
 The spin-independent part $J_0$ does not contribute to current in the linear order in $B_x$ so $J=J_1$. Defining $D\equiv E_2^2-E_1^2=4\sqrt{B^2(\epsilon_k^2+\Delta^2)+\epsilon_k^2\alpha^2k^2}$, Eq.~(\ref{acu4}) coincides with the current formula $(\ref{cu4})$ in the main text.

\subsection{Magnetoelectric effect for perpendicular fields and adiabatic pumping}

As discussed in the main text, the perpendicular magnetoelectric effect (for fields with $B_x=0$) and adiabatic pumping is determined by the quantity
\begin{align}\label{omegapp}
\Omega_{xy}(k)=\left(-\frac{i\alpha^2 }{2}\right)\sum_{n,n'}\frac{\mathrm{Tr}\left[\sigma_x P_{n'}\sigma_y P_n\right] }{(E_n-E_{n'})^2}(f_n-f_{n'}),
\end{align}
where $P_n$ is a projection operator to the eigenstate of the Bogoliubov- de Gennes Hamiltonian with eigenvalue $E_n$ given by Eq.~(\ref{E}) in the main text and $f_n$ is the Fermi function at energy $E_n$. The projection operators are given by $P_i=\prod_{j\neq i}\frac{H-E_j}{E_i-E_j}$ which in this case simplify to $P_{\pm 1}=\frac{1}{2}\left(1\pm\frac{H}{E_1}\right)\frac{H^2-E_2^2}{E_1^2-E_2^2}$ and analogously for $P_{\pm 2}$ by interchanging subscripts $1 \leftrightarrow 2$. Evaluation of the traces in Eq.~(\ref{omegapp}) is now in principle straightforward but a tedious task. The explicit form of $\Omega_{xy}$ is given by
\begin{align}\label{q3}
&\Omega_{xy}(k)=\frac{4\alpha^2 B}{E_1E_2} \left[\left(-\frac{\Delta^2+\epsilon_k^2}{E_1+E_2}+\frac{E_1+E_2}{4} \right) \frac{\left(f_1+f_2-1\right)}{(E_1+E_2)^2}+
\left(\frac{\Delta^2+\epsilon_k^2}{E_2-E_1}-\frac{E_2-E_1}{4} \right) \frac{\left(f_1-f_2 \right) }{(E_2-E_1)^2} \right].
\end{align}
This form shows that $\Omega_{xy}$ vanishes if either $B$ or $\alpha$ vanishes. In the linear response current (Eq.~(\ref{lin3}) in the main text) $B$ plays the role of the static part of the field while in the adiabatic pumping process (Eq.~(\ref{q2}) in the main text) $B=B(\lambda)$ is the slowly varying function of the adiabatic parameter. In the adiabatic pumping process the energies depend on the adiabatic parameter through $B(\lambda)$.

It is worth noting that in Eq.~(\ref{omegapp}) the terms in the sum corresponding to pairs $n=1/2$, $n'=-1/-2$ vanish, since the system possess a selection rule $\langle E_n(k) |\sigma_{y/z}|E_{-n}(k)\rangle=0$. This selection rule can be shown with the help of the chiral symmetry $\{H,\sigma_x\tau_y\}=0$ (which is present when $B_x=0$). Therefore perpendicular magnetic fields cannot couple the $E_1$ and $E_{-1}$ bands and $\Omega_{xy}$ remains analytic at the phase transition where $E_{-1}(k)=E_{1}(k)=0$ at $k=0$.

\end{document}